\documentclass[aip,apl,reprint]{revtex4-1}

\usepackage{graphicx}% Include figure files
\usepackage{amsmath}
\usepackage{textcomp} %texterweiterungen
\usepackage{epstopdf}

\usepackage{color}

\begin{document}

\title{Frequency shift keying by current modulation in a MTJ-based STNO with high data rate}
%\title{High data rate in frequency shift keying by current modulation in\textcolor[rgb]{1,0,0}{/of} a MTJ-based STNO}

\author{A. Ruiz-Calaforra}
\email{ana.ruizcala@gmail.com}
\affiliation{Univ. Grenoble Alpes, CEA, CNRS, Grenoble INP\thanks{Institute of Engineering Univ. Grenoble Alpes}, INAC, SPINTEC, F-38000 Grenoble, France}
%\affiliation{Univ. Grenoble Alpes, INAC-SPINTEC, F-38000 Grenoble, France}
%\affiliation{CEA, INAC-SPINTEC, F-38000 Grenoble, France}
%\affiliation{CNRS, SPINTEC, F-38000 Grenoble, France}

\author{A. Purbawati}
\altaffiliation{A.Ruiz-Calaforra and A. Purbawati contributed equally to this work.}
\affiliation{Univ. Grenoble Alpes, CEA, CNRS, Grenoble INP\thanks{Institute of Engineering Univ. Grenoble Alpes}, INAC, SPINTEC, F-38000 Grenoble, France}

\author{T. Br\"acher}
\affiliation{Univ. Grenoble Alpes, CEA, CNRS, Grenoble INP\thanks{Institute of Engineering Univ. Grenoble Alpes}, INAC, SPINTEC, F-38000 Grenoble, France}

\author{J. Hem}
\affiliation{Univ. Grenoble Alpes, CEA, CNRS, Grenoble INP\thanks{Institute of Engineering Univ. Grenoble Alpes}, INAC, SPINTEC, F-38000 Grenoble, France}
%Jérôme Hem

\author{C. Murapaka}
%Chandrasekhar Murapaka
\affiliation{Univ. Grenoble Alpes, CEA, CNRS, Grenoble INP\thanks{Institute of Engineering Univ. Grenoble Alpes}, INAC, SPINTEC, F-38000 Grenoble, France}

\author{E. Jim\'{e}nez}
%Erika Jiménez Romero
\affiliation{Univ. Grenoble Alpes, CEA, CNRS, Grenoble INP\thanks{Institute of Engineering Univ. Grenoble Alpes}, INAC, SPINTEC, F-38000 Grenoble, France}

\author{D. Mauri}
%Daniele Mauri
\affiliation{Western Digital, 5600 Great Oaks Parkway, San Jose, CA 95119}

\author{A. Zeltser}
%Alexander Zeltser
\affiliation{Western Digital, 5600 Great Oaks Parkway, San Jose, CA 95119}

\author{J.~A. Katine}
%Jordan Katine
\affiliation{Western Digital, 5600 Great Oaks Parkway, San Jose, CA 95119}

\author{M.-C. Cyrille}
%Marie-Claire Cyrille
\affiliation{Univ. Grenoble Alpes, F-38000 Grenoble, France}
\affiliation{CEA-LETI MINATEC-CAMPUS, 17 F-38054 Grenoble, France}

\author{L.~D. Buda-Prejbeanu}
\affiliation{Univ. Grenoble Alpes, CEA, CNRS, Grenoble INP\thanks{Institute of Engineering Univ. Grenoble Alpes}, INAC, SPINTEC, F-38000 Grenoble, France}

\author{U. Ebels}
\affiliation{Univ. Grenoble Alpes, CEA, CNRS, Grenoble INP\thanks{Institute of Engineering Univ. Grenoble Alpes}, INAC, SPINTEC, F-38000 Grenoble, France}

%\date{\today}

\begin{abstract}

Spin torque nano-oscillators are nanoscopic microwave frequency generators which excel due to their large frequency tuning range and agility for amplitude and frequency modulation. Due to their compactness, they are regarded as suitable candidates for applications in wireless communications, where cost-effective and CMOS-compatible standalone devices are required. In this work, we study the ability of a magnetic-tunnel-junction (MTJ) based spin torque nano-oscillator to respond to a binary input sequence encoded in a square-shaped current pulse for its application as a frequency-shift-keying (FSK) based emitter. We demonstrate that below the limit imposed by the spin torque nano-oscillators intrinsic relaxation frequency, an agile variation between discrete oscillator states is possible. For this kind of devices, we demonstrate FSK up to data rates of $400\,\mathrm{Mbps}$ which is well suited for the application of such oscillators in wireless networks.

\end{abstract}


\maketitle

%%%%%%%%%%%%%%%   Introduction %%%%%%%%%%%%%%%%%%

Future wireless communications demand a next generation of rf devices with low power consumption, which are compact and multifunctional. Spin torque nano-oscillators (STNOs) use the spin transfer torque (STT) effect in thin magnetic nanoelements to induce steady state oscillations via spin momentum transfer from the conduction electrons to the local magnetization\cite{Slonzewski1996JMMM, Berger1996PRB, Tsoi1998PRL, Kiselev2003Nature}. These magnetic oscillations are then converted into an oscillatory magnetoresistive output signal at frequencies from $100\,\mathrm{MHz}$ to several tens of $\mathrm{GHz}$ depending on the STNO type\cite{Russek-Book-2010,Villard-2010,Chen2016IEEE}. STNOs have been identified as a promising class of rf devices for applications in wireless communication as well as read heads\cite{Zeng-Nanoscale-2013, Mizushima-JAP-2010, Braganca-2010-Nano}. This is due to their sustained microwave frequency oscillations and their large frequency tunability by the injected dc current and the applied field, as well as their large agility which allows for a fast modulation of the STNO output\cite{Pufall2005APL,Muduli2010PRB,Quinsat2014APL}. In addition, STNOs are highly compact due to their nanoscale dimension and compatible with with complementary metal-oxide semiconductor (CMOS).

One of the key features of STNOs is the tunability of their amplitude and frequency by a modulation of the applied current or magnetic field\cite{Quinsat2014APL,Anike2016APL}. This can be used to encode digital information via amplitude shift keying (ASK)\cite{Choi2014SciRep, Oh2014IEEEMWCL, Sharma2015IEEETransMagn} or frequency shift keying (FSK)\cite{Manfrini2009APL,Manfrini2011JAP}. In these concepts, the amplitude or the frequency of the STNO output voltage is shifted between two discrete values corresponding to digital values of “0” and “1”. For instance, first results for STNO FSK operation under a pulsed variation of the magnetic field have been reported and modulation bit rates in the order of Gbps have been achieved\cite{Nagasawa2011JAP, Nagasawa2012JAP}. While this concept is particularly suited for read-head applications, the use of a modulated current is favorable for applications in wireless communications, since it allows for a particularly simple and compact device design which can be easily integrated. Digital current modulation has been reported in the context of amplitude On-Off Keying (OOK)\cite{Choi2014SciRep, Oh2014IEEEMWCL, Sharma2015IEEETransMagn}, demonstrating communication over $1\,\mathrm{m}$ distance with data rates of $0.2-2\,\mathrm{Mbps}$, as well in the context of FSK using a vortex magnetic tunnel junction (MTJ)-based STNO achieving modulation frequencies up to $10\,\mathrm{MHz}$ (data rates of $20\,\mathrm{Mbps}$)\cite{Manfrini2011JAP}. More recently, the implementation of a homogeneously in-plane magnetized MTJ-based STNO for FSK on a printed circuit board with data rates up to $20\,\mathrm{Mbps}$ was demonstrated\cite{Rui2017}.

Here, we report on the study of current-modulated FSK using the same MTJ-based STNO as Ref. \onlinecite{Rui2017}. We discuss the device characteristics of the free-running oscillator and determine the limits of the STNO to respond to a change in the applied current. Consequently, the STNOs ability to perform FSK is demonstrated experimentally, focusing on the maximum achievable data rate. We demonstrate that at least up to the relaxation frequency $f_\mathrm{p}$, FSK of the STNO is possible since the STNO frequency follows the modulation current, and data rates up to $400\,\mathrm{Mbps}$ could be achieved.  

%\section{Samples and experiments}
%\label{experiment}
%
%\begin{figure*}[t!]
%	  \begin{center}
%    \scalebox{1}{\includegraphics[width=1\textwidth, clip]{Fig1-horizontal.eps}}
%    \end{center}
%	  \caption{\label{fig:Fig1}(a) Power spectral density (PSD) (color-coded) as a function of frequency and field for a fixed current of $I_\mathrm{dc}=1\,\mathrm{mA}$. (b) Frequency, (c) linewidth and integrated power vs current $I_\mathrm{dc}$ for an applied field of $H=1\,\mathrm{kOe}$. (d) Snapshot of the time resolved amplified output voltage of the STNO for $I_\mathrm{dc}=1.05\,\mathrm{mA}$ and (e) corresponding amplitude autocorrelation $\kappa_p(t)$ and exponential fit.}
%\end{figure*}
%%%%%%%%%
Experiments have been conducted on a homogeneously in-plane magnetized MTJ nanopillar with the following stack composition (thickness in nanometer) IrMn (6.1)/CoFe(1.8)/Ru(0.4)/CoFeB(2)/MgO/CoFe(0.5)\\/CoFeB(3.4) and a nominal resistance area (RA) product of $1\,\mathrm{\Omega\mu m^2}$\cite{Houssameddine2008APL, Houssameddine2009PRL, Quinsat2010APL}. The nanopillar, with size $65\times130\,\mathrm{nm^2}$, exhibits a resistance of $228\,\mathrm{\Omega}$ in the antiparallel configuration and a tunnel magnetoresistance (TMR) value of $46\,\%$. The STNO is first characterized in terms of its free-running characteristics to evaluate its performance and tunability by the applied charge current.

This dynamic characterization of the excitation of the in-plane precession (IPP) of the magnetization of the free layer has been performed using a standard radiofrequency setup. The output voltage signal upon injection of a DC current into the STNO is first amplified by a $43\,\mathrm{dB}$ amplifier and then passed through a high pass filter ($>3\,\mathrm{GHz}$). The signal is split and analyzed via a spectrum analyzer in the frequency domain and via a fast, real time oscilloscope in the time domain. From the frequency-domain measurement, the free running frequency $f$, the full width at half maximum (FWHM) linewidth $\Delta f$ and, by dividing out the overall amplification of the microwave circuit, the integrated peak power $P$ of the STNO are determined as a function of the external control parameters dc current $I_\mathrm{dc}$ and magnetic field $H$. 
%%%%%%%%
\begin{figure}[t]
	  \begin{center}
    \scalebox{1}{\includegraphics[width=1\linewidth, clip]{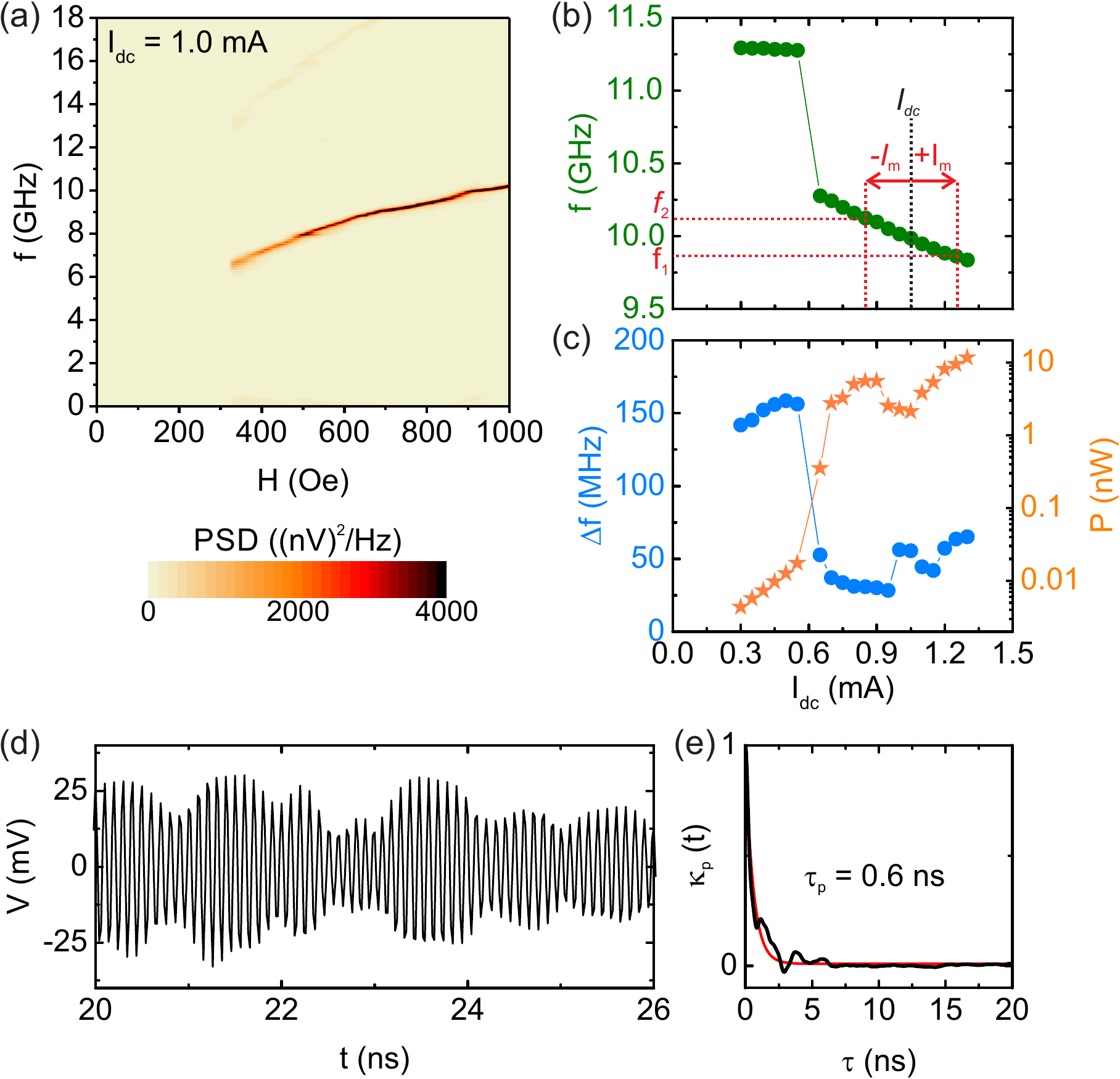}}
    \end{center}
	  \caption{\label{fig:Fig1}(a) Power spectral density (PSD) (color-coded) as a function of frequency and field for a fixed current of $I_\mathrm{dc}=1\,\mathrm{mA}$. (b) Frequency, (c) linewidth and integrated power vs current $I_\mathrm{dc}$ for an applied field of $H=1\,\mathrm{kOe}$. (d) Snapshot of the time resolved amplified output voltage of the STNO for $I_\mathrm{dc}=1.05\,\mathrm{mA}$ and (e) corresponding amplitude autocorrelation $\kappa_p(t)$ and exponential fit.}
\end{figure}
%%%%%%%%%

Figure\,\ref{fig:Fig1}\,(a) shows the microwave emission spectrum of the free layer in the antiparallel configuration as a function of the applied field, which features an in-plane angle of $15\,\mathrm{^\circ}$ with respect to the long axis of the elliptically shaped nanopillar, for a fixed current of $I_\mathrm{dc}=1\,\mathrm{mA}$. A single mode Kittel-like excitation, with frequencies between $6\,\mathrm{GHz}$ to $10\,\mathrm{GHz}$ for the measured field range is observed. A weak second harmonic above $12\,\mathrm{GHz}$ can also be detected. Figures\,\ref{fig:Fig1}\,(b) and (c) show the resulting $f$, $\Delta f$ and $P$ for the investigated STNO as a function of $I_\mathrm{dc}$ for a fixed field $H=1\,\mathrm{kOe}$. The threshold current $I_\mathrm{C}$, given by the jump in $f$ of about $1\,\mathrm{GHz}$ and the large decrease in $\Delta f$ accompanied by a strong increase of the emitted power, is estimated to be $I_\mathrm{C}\approx0.65\,\mathrm{mA}$. Above this threshold current, the microwave excitations are steady state auto-oscillations. The steady state excitation features frequencies above $9.8\,\mathrm{GHz}$ and linewidths lower than $60\,\mathrm{MHz}$ with output powers up to $10\,\mathrm{nW}$. The frequency-current tuning $\mathrm{d}f/\mathrm{d}I$ is linear and negative (frequency redshift) in the range of $I_\mathrm{dc} =0.65\,\mathrm{mA}$ to $1.3\,\mathrm{mA}$ with a value of $\mathrm{d}f/\mathrm{d}I=-677\,\mathrm{MHz/mA}$.

Further insight into the behavior of the STNO in the steady state can be inferred from a real-time observation of its output voltage. For such a time domain study, a high speed single-shot oscilloscope with a $20\,\mathrm{GHz}$ bandwidth and a $50\,\mathrm{GSa/s}$ sampling rate is used. $40\,\mathrm{\mu s}$ long time traces have been acquired in the current range of the steady state regime with a time resolution of $20\,\mathrm{ps}$. In order to eliminate low frequency noise and higher harmonic contributions, a $4\,\mathrm{GHz}$ numerical band pass filter centered on the first harmonic is applied to the time-domain data. Figure\,\ref{fig:Fig1}\,(d) illustrates\footnote{All shown time traces in the manuscript have been smoothed using a Savitzky-Golay filter with a $2.3\,\mathrm{ns}$ window.} a $6\,\mathrm{ns}$ segment of an output voltage time trace after filtering for an applied current $I_\mathrm{dc} =1.05\,\mathrm{mA}$. The presented segment is representative for the temporal behavior of the STNO, which exhibits a visible amplitude fluctuation but overall presents a stable steady-state oscillation with a very small number of extinctions\cite{Houssameddine2009PRL}.

To analyze the time traces, the standard evaluation based on the Hilbert transform is employed\cite{Quinsat2010APL, Tiberkevich2008PRB, Slavin2009IEEETransMagn, Bianchini2010APL, Picinbono-1997-IEEETransSignalProcess}. From the Hilbert transform, the amplitude envelope $p(t)$ and fluctuation $\delta p(t)$ as well as the time-varying phase $\Phi(t)$ are extracted. The instantaneous frequency $f_i(t)$ is obtained from the time-varying phase via $f_i(t)=1/(2\pi)\mathrm{d}\Phi/\mathrm{d}t$. The amplitude fluctuation is used to estimate the amplitude relaxation time $\tau_\mathrm{p}$ following the approach discussed in Refs.~\onlinecite{Quinsat2010APL, Tiberkevich2008PRB, Slavin2009IEEETransMagn, Bianchini2010APL}. The amplitude relaxation time corresponds to the time the oscillator needs to respond to a change of the oscillation state. In the context of frequency shift keying by current modulation, it determines the highest achievable data rate\cite{Quinsat2014APL, Anike2016APL}, which is given by two times the relaxation frequency $f_\mathrm{p} = (2\pi \tau_\mathrm{p})^{-1}$. $\tau_p$ is determined from an exponential fit of the autocorrelation $\kappa_p(t)= \left\langle \delta p(t)\delta p(0)\right\rangle$. Figure\,\ref{fig:Fig1}\,(e) exemplarily shows the autocorrelation $\kappa_p(t)$ together with an exponential fit for a current of $I_\mathrm{dc}=1.05\,\mathrm{mA}$. From the fit, $\tau_\mathrm{p}=0.6\,\mathrm{ns}$ is obtained. Such fits have been performed throughout the current range of the steady state regime. All obtained values of $\tau_\mathrm{p}$ are on the order of $0.6\,\mathrm{ns}$, with a slight increase with increasing $I_\mathrm{dc}$. This corresponds to a relaxation frequency of $f_\mathrm{p}\approx260\,\mathrm{MHz}$, which imposes an upper limit of the data rate ($2f_\mathrm{p}$) of around $500\,\mathrm{Mbps}$.

%\section{Frequency Shift Keying}
%\label{FSK}
%
%%%%%%%%
\begin{figure*}[t]
	  \begin{center}
    \scalebox{1}{\includegraphics[width=0.9\textwidth, clip]{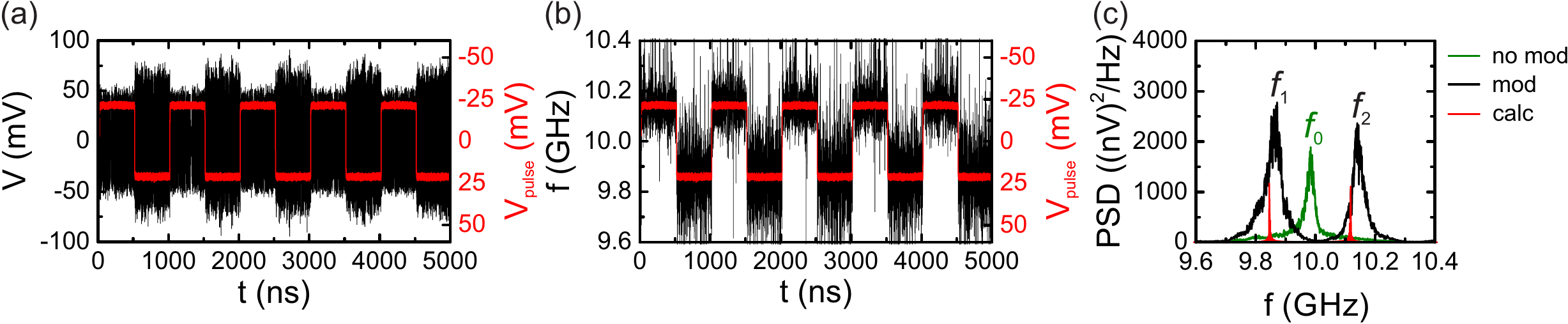}}
    \end{center}
	  \caption{\label{fig:Fig2}a) Snapshot from the amplified output voltage of the modulated STNO with $f_\mathrm{m}=1\,\mathrm{MHz}$ (black) together with the injected current (red). b) Corresponding instantaneous frequency obtained from the Hilbert transformation. Please note that the voltage axis of the pulses is inverted for better comparison with the frequency shift due to the negative $\mathrm{d}f/\mathrm{d}I$. c) Corresponding power spectral density (PSD) (black) together with the PSD of the free-running STNO (green). The red lines in (c) mark the calculated position of the peaks of the modulated signal (see text).}
\end{figure*}
%%%%%%%%%

After this characterization of the free-running STNO, in the following, we will focus on the experimental realization of FSK. For this experiment, we choose a free running carrier frequency $f_0 \approx 9.98\,\mathrm{GHz}$ ($I_\mathrm{dc}=1.05\,\mathrm{mA}$) which is well situated in the steady state regime (see Fig.\,\ref{fig:Fig1}\,(b)). The constant current needed to achieve steady state oscillations at $f_0$ is applied by a sourcemeter. The current is modulated around this central value by using an arbitrary waveform generator, which generates a series of square shaped pulses resulting in a voltage modulation with a $50\,\%$ duty cycle and a rise and fall time of $1\,\mathrm{ns}$ after passing through the bias-tee. Hence, during half of the pulse period $T_\mathrm{m}=1/f_\mathrm{m}$, where $f_\mathrm{m}$ is the modulation frequency, the STNO is excited by a current $I_\mathrm{dc}+I_\mathrm{m}$ and oscillates at a frequency $f_1$ and during the second half it is excited by a current $I_\mathrm{dc}-I_\mathrm{m}$ and oscillates at a frequency $f_2$. The peak-to-peak modulation corresponds to $2I_\mathrm{m}\approx 0.4\,\mathrm{mA}$. The modulation frequency $f_\mathrm{m}$ was varied from $1\,\mathrm{MHz}$ to $200\,\mathrm{MHz}$. For further analysis, the modulated signal is acquired using a fast oscilloscope. Figure\,\ref{fig:Fig2}\,(a) shows a snapshot of the modulated voltage signal for $f_\mathrm{m}=1\,\mathrm{MHz}$. The red curve corresponds to the injected pulsed current. As can be seen, the variation of the current gives rise to a weak amplitude modulation, which could be employed to perform amplitude shift keying (ASK). To study and visualize the FSK, the instantaneous frequency as a function of time is extracted via the Hilbert transform as outlined above. The result can be seen in Fig.\,\ref{fig:Fig2}\,(b), while Fig.\,\ref{fig:Fig2}\,(c) shows the corresponding power spectral density (PSD) of the modulated signal (mod) and for the non-modulated signal (no mod), which corresponds to the free running oscillation at $I_\mathrm{dc}=1.05\,\mathrm{mA}$ (c.f Fig.\,\ref{fig:Fig1}\,(d)). From the instantaneous frequency and the PSD, it can be seen that the oscillation frequency of the STNO is modulated between two levels ($f_1=9.84\,\mathrm{GHz}$ and $f_2=10.12\,\mathrm{GHz}$) demonstrating FSK for $f_\mathrm{m}=1\,\mathrm{MHz}$ with a frequency splitting $\delta f=280\,\mathrm{MHz}$. The two frequency levels $f_1$ and $f_2$ feature a visible phase noise, which is responsible for the finite linewidth of the oscillator\cite{Houssameddine2009PRL}. As can be seen from Figs.\,\ref{fig:Fig2}\,(b) and (c), the frequency splitting between the two states is well above the linewidth and, thus, the two states can be clearly distinguished despite the finite noise.

%%%%%%%%
\begin{figure}[b]
	  \begin{center}
    \scalebox{1}{\includegraphics[width=1\linewidth, clip]{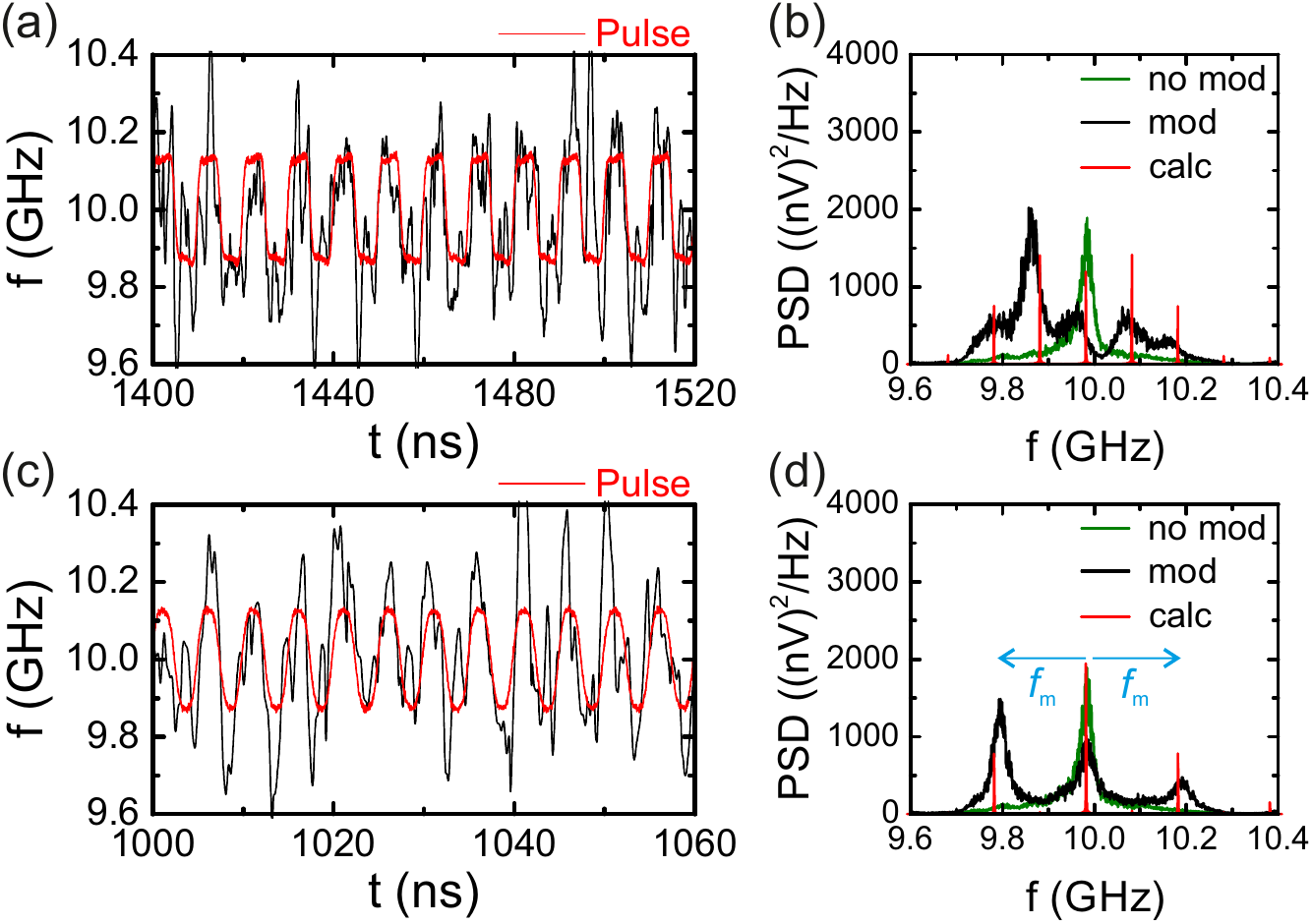}}
    \end{center}
	  \caption{\label{fig:Fig3} Instantaneous frequency of the modulated STNO and corresponding power spectral density (PSD) (black) for $f_\mathrm{m} = 100\,\mathrm{MHz}$ ((a) and (b), respectively) and $f_\mathrm{m} = 200\,\mathrm{MHz}$ ((c) and (d)). For comparison, the PSD of the free-running STNO is included in (b) and (d) (green).}
\end{figure}
%%%%%%%%%
Figure\,\ref{fig:Fig3} shows the instantaneous frequency obtained for higher data rates of $f_\mathrm{m} = 100\,\mathrm{MHz}$ and $f_\mathrm{m} = 200\,\mathrm{MHz}$. It can be seen that the frequency follows the modulation signal even for such high modulation frequencies. Figure\,\ref{fig:Fig3} shows the experimental limitation of our setup: For $f_\mathrm{m} = 200\,\mathrm{MHz}$, the pulses themselves become practically sinusoidal due to their finite rise time of $1\,\mathrm{ns}$. This corresponds to a continuous variation of the instantaneous frequency rather than a discrete variation between two discrete states. Nevertheless, since this modulation frequency is still well below the relaxation frequency $f_\mathrm{p}$, and since the rise and fall time of the pulses is larger than the relaxation time $\tau_\mathrm{p}$, the oscillator can readily follow the modulation. When comparing  the PSD results for the higher data rates (Fig.\,\ref{fig:Fig3}\,(b) and (d)) with the one from $f_\mathrm{m}=1\,\mathrm{MHz}$ (Fig.\,\ref{fig:Fig2}\,(c)), one can see that this behavior is reflected in the PSD, which continuously evolves from two discrete peaks to a central peak with two satellite peaks at a distance of $f_\mathrm{m}$. 

This evolution of the PSD can be well understood by the variation of the modulation frequency with respect to the frequency splitting between the two oscillator states. To do so, it is instructive to look at the PSD expected from a perfect, noise-free oscillator which instantaneously follows the modulation signal. For such an oscillator, the time-dependent output voltage is given by:
\begin{align}
\label{Eq:Vout}
V(t) = \frac{V_0}{2}\mathrm{sin}\left(2\pi\cdot f_0 \cdot t + \pi \cdot \delta f \int_0^t{V_\mathrm{m}(t')\mathrm{d}t'}\right).
\end{align}

Here, $V_0$ is the peak to peak voltage. $V_\mathrm{m}$ is the normalized modulation voltage (from -1 to +1, including the conversion between applied voltage and frequency shift), which we obtain from the experimentally used pulses depicted by the red curves in Figs.\,\ref{fig:Fig2}\,(a) and (b) and \ref{fig:Fig3}\,(a) and (c). The PSD obtained by a Fourier transformation of this modulated voltage signal is shown by the red lines in Figs.\,\ref{fig:Fig2}\,(c),\ref{fig:Fig3}\,(b) and \ref{fig:Fig3}\,(d). As can be seen from these figures, there is a good correspondence between the expected peak position and splitting between the peaks, where minor deviations are mainly due to temperature and field drifts during the experiment. Thus, also from the PSD, it can be concluded that the oscillator follows the modulated voltage sufficiently fast since $f_\mathrm{m}<f_\mathrm{p}$.

In summary, we have demonstrated frequency shift keying with data rates up to $400\,\mathrm{Mbps}$ with homogeneously in-plane magnetized MTJ-based STNOs at a carrier frequency of $9.98\,\mathrm{GHz}$. The highest used modulation frequency $200\,\mathrm{MHz}$ is below the STNOs relaxation frequency $f_\mathrm{p}$. Thus, the STNO is able to follow the frequency change quasi-instantaneously. The main limitations in our experiment arise from the finite rise times of the used pulse pattern and the phase-noise of the STNO. Our results demonstrate that the high relaxation frequency of homogeneously in-plane magnetized MTJ-based STNOs readily allows for data rates on the order of several $100\,\mathrm{Mbps}$ by current modulation. Thus, they are well suited for their application as compact standalone nodes in wireless communication networks.

\section*{Acknowledgments}

This work was supported the EC FP7 MOSAIC N° 317950 and MAGICAL.

\end{document}